\documentclass[12pt]{article}

\evensidemargin \oddsidemargin
\usepackage[bottom]{footmisc}
\usepackage[affil-it]{authblk}

\usepackage{cite,amsmath,epsfig,amssymb,graphicx,cite,float}

\newcommand{\brat}{{\rm BR}}

\newcommand{\be}{\begin{equation}}
\newcommand{\ee}{\end{equation}}
\newcommand{\bea}{\begin{eqnarray}}
\newcommand{\eea}{\end{eqnarray}}

\newcommand{\complex}{{{\rm I} \kern -.59em {\rm C}}}

\def\beq{\begin{equation}}
\def\eeq{\end{equation}}

\newcommand{\FUTB}{{\bf FUT}}
\def\reffi#1{\mbox{Fig.~\ref{#1}}}

\newcommand{\Mh}{M_h}

\newcommand{\mt}{m_{t}}

\newcommand{\mb}{m_{b}}

\newcommand{\tsf}{\theta\kern-.20em_{\tilde{f}}}
\newcommand{\tsfp}{\theta\kern-.20em_{\tilde{f}\prime}}
\newcommand{\tsq}{\theta\kern-.15em_{\tilde{q}}}

\newcommand{\VL}{\left( \begin{array}{c}}
\newcommand{\VR}{\end{array} \right)}
\newcommand{\ML}{\left( \begin{array}{cc}}
\newcommand{\MLd}{\left( \begin{array}{ccc}}
\newcommand{\MLv}{\left( \begin{array}{cccc}}
\newcommand{\MR}{\end{array} \right)}

\newcommand{\tev}{\,\, \mathrm{TeV}}
\newcommand{\gev}{\,\, \mathrm{GeV}}

\newcommand{\BC}{\begin{center}}
\newcommand{\EC}{\end{center}}
\newcommand{\BE}{\begin{equation}}
\newcommand{\EE}{\end{equation}}
\newcommand{\BEA}{\begin{eqnarray}}
\newcommand{\BEAnn}{\begin{eqnarray*}}
\newcommand{\EEA}{\end{eqnarray}}
\newcommand{\EEAnn}{\end{eqnarray*}}

\newcommand{\id}{{\rm 1\kern-.12em
\rule{0.3pt}{1.5ex}\raisebox{0.0ex}{\rule{0.1em}{0.3pt}}}}
\newcommand{\lsim}
{\;\raisebox{-.3em}{$\stackrel{\displaystyle <}{\sim}$}\;}

\newcommand{\br}{{\rm BR}}

\begin{document}

\title{Reduction of Couplings  in Quantum Field Theories with applications in  Finite Theories and the MSSM}
\date{}
\author{S. Heinemeyer$^1$\thanks{email: Sven.Heinemeyer@cern.ch}, M. Mondrag\'on$^2$\thanks{email: myriam@fisica.unam.mx}, N. Tracas$^3$\thanks{email: ntrac@central.ntua.gr} and G. Zoupanos$^{3,4}$\thanks{email: George.Zoupanos@cern.ch}}

\affil{\small $^1$Instituto de F\'{\i}sica de Cantabria
  (CSIC-UC),  E-39005 Santander, Spain, \\
$^2$Instituto de F\'{\i}sica,  Universidad Nacional
Aut\'onoma de M\'exico,  A.P. 20-364, M\'exico 01000\\ %
$^3$Physics Deptartment,   Nat. Technical University, 
 157 80 Zografou, Athens, Greece\\%
$^4$Institut f\"ur Theoretische Physik,
Universit\"at Heidelberg, 
Philosophenweg 16, D-69120 Heidelberg, Germany} %
\maketitle

\abstract{We apply the method of reduction of couplings in a Finite
  Unified Theory and in the MSSM. The method consists on searching for
  renormalization group invariant relations among couplings of a
  renormalizable theory holding to all orders in perturbation
  theory. It has a remarkable predictive power since, at the
  unification scale, it leads to relations between gauge and Yukawa
  couplings in the dimensionless sectors and relations involving the
  trilinear terms and the Yukawa couplings, as well as a sum rule
  among the scalar masses and the unified gaugino mass in the soft
  breaking sector.  In both the MSSM and the FUT model we predict the
  masses of the top and bottom quarks and the light Higgs in
  remarkable agreement with the experiment. Furthermore we also 
  predict the masses of the other Higgses, as well as the
  supersymmetric spectrum, both being in very confortable agreement
  with the LHC bounds on Higgs and supersymmetric particles.  }

\section{Introduction}

The discovery of a Higgs boson
\cite{Aad:2012tfa,ATLAS:2013mma,Chatrchyan:2012ufa,Chatrchyan:2013lba}
at the LHC completes the search
for the particles of the Standard Model (SM), and confirms the 
existence of a Higgs field and the 
spontaneous electroweak symmetry breaking mechanism as the way to
explain the masses of the fundamental particles.  The over twenty free parameters of
the SM, the hierarchy problem, the existence of Dark Matter, 
the very
small masses of the neutrinos, among others, point towards a more
fundamental theory, whose goal among others should be to explain at least some of these facts.

The main achievement expected from a unified description of
interactions is to understand the
large number of free parameters of the Standard Model (SM)
in terms of a few fundamental ones. In other words, to achieve {\it
  reduction of couplings} at a more fundamental level. 
To reduce the number of free parameters of a theory, and thus render
it more predictive, one is usually led to introduce more symmetry.
Supersymmetric Grand
Unified Theories (GUTs) are very good examples of such a procedure
\cite{Pati:1973rp,Georgi:1974sy,Georgi:1974yf,Fritzsch:1974nn,Carlson:1975gu,Dimopoulos:1981zb,Sakai:1981gr}.

For instance, in the case of minimal $SU(5)$, because of (approximate)
gauge coupling unification, it was possible to reduce the gauge
couplings to one.
 LEP data \cite{Amaldi:1991cn} 
seem to
suggest that a further symmetry, namely $N=1$ global supersymmetry
\cite{Dimopoulos:1981zb,Sakai:1981gr} 
should also be required to make the prediction viable.
GUTs can also
relate the Yukawa couplings among themselves, again $SU(5)$ provided
an example of this by predicting the ratio $M_{\tau}/M_b$
\cite{Buras:1977yy} in the SM.  Unfortunately, requiring
more gauge symmetry does not seem to help, since additional
complications are introduced due to new degrees of freedom and  in the
ways and channels of breaking the symmetry.

A natural extension of the GUT idea is to find a way to relate the
gauge and Yukawa sectors of a theory, that is to achieve Gauge-Yukawa
Unification (GYU) \cite{Kubo:1995cg,Kubo:1997fi,Kobayashi:1999pn}.  A
symmetry which naturally relates the two sectors is supersymmetry, in
particular $N=2$ supersymmetry \cite{Fayet:1978ig}.  It turns out, however, that $N=2$
supersymmetric theories have serious phenomenological problems due to
light mirror fermions.  Also in superstring theories and in composite
models there exist relations among the gauge and Yukawa couplings, but
both kind of theories have other phenomenological problems, which we are not
going to address here.

A complementary strategy in searching for a more fundamental theory,
consists in looking for all-loop renormalization group invariant (RGI)
relations \cite{Zimmermann:1984sx,Oehme:1984yy} holding below the
Planck scale, which in turn are preserved down to the unification
scale
\cite{Kapetanakis:1992vx,Mondragon:1993tw,Kubo:1994bj,Kubo:1994xa,Kubo:1995hm,Kubo:1995cg,Kubo:1996js,Kubo:1997fi,Kobayashi:1999pn}. 
Through this method of reduction of couplings
\cite{Zimmermann:1984sx,Oehme:1984yy} it is possible
to achieve
Gauge-Yukawa Unification 
\cite{Kubo:1995cg,Kubo:1995zg,Kobayashi:1999pn,Kubo:1997fi}.
Even more remarkable is the fact that it is possible to find RGI
relations among couplings that guarantee finiteness to all-orders in
perturbation theory 
\cite{Lucchesi:1987he,Piguet:1986td,Lucchesi:1996ir}.

Although supersymmetry seems to be an essential feature for a
successful realization of the above programme, its breaking has to be
understood too, since it has the ambition to supply the SM with
predictions for several of its free parameters. Indeed, the search for
RGI relations has been extended to the soft supersymmetry breaking
sector (SSB) of these theories \cite{Kubo:1996js,Jack:1995gm,Zimmermann:2001pq}, which
involves parameters of dimension one and two.

\section{The Method of Reduction of Couplings}

In this section we will briefly outline the reduction of couplings method. Any RGI
relation among couplings (i.e. which does not depend on the
renormalization scale $\mu$ explicitly) can be expressed, in the
implicit form $\Phi (g_1,\cdots,g_A) ~=~\mbox{const.}$, which has to
satisfy the partial differential equation (PDE)
\be
\frac{d \Phi}{dt}=
\sum_{a=1}^{A}\,\frac{\partial \Phi}{\partial g_{a}}
\frac{dg_a}{dt}=
\sum_{a=1}^{A} \,\frac{\partial \Phi}{\partial g_{a}}\beta_{a}=
{\vec \nabla} \Phi \cdot {\vec \beta} =0,
\ee
where $t=\ln\mu$  and $\beta_a$ is the $\beta$-function of $g_a$. This PDE is
equivalent to a set of ordinary differential equations, the so-called
reduction equations (REs)
\cite{Zimmermann:1984sx,Oehme:1984yy,Oehme:1985jy},
\be \beta_{g} \,\frac{d
g_{a}}{d g} =\beta_{a}~,~a=1,\cdots,A~,
\label{redeq}
\ee
where $g$ and $\beta_{g}$ are the primary coupling and its
$\beta$-function, and the counting on $a$ does not include $g$. Since
maximally ($A-1$) independent RGI ``constraints'' in the
$A$-dimensional space of couplings can be imposed by the $\Phi_a$'s,
one could in principle express all the couplings in terms of a single
coupling $g$.  The strongest requirement in the search for RGI relations is to demand power series
solutions to the REs,
\be g_{a} = \sum_{n=0}\rho_{a}^{(n)}\,g^{2n+1}~,
\label{powerser}
\ee
which formally preserve perturbative renormalizability. Remarkably,
the uniqueness of such power series solutions can be decided already
at the one-loop level
\cite{Zimmermann:1984sx,Oehme:1984yy,Oehme:1985jy}.

Searching for a power series solution of the form (\ref{powerser}) to
the REs (\ref{redeq}) is justified  since various
couplings in supersymmetric theories have the same asymptotic
behaviour, thus one can rely that keeping only the first terms in the expansion is a
good approximation in realistic applications.

\section{Reduction of Couplings in Soft Breaking Terms }

The method of reducing the dimensionless couplings was 
extended\cite{Kubo:1996js,Jack:1995gm,Zimmermann:2001pq} to the soft
supersymmetry breaking (SSB) dimensionful parameters of $N = 1$
supersymmetric theories.  In addition it was found
\cite{Kawamura:1997cw,Kobayashi:1997qx} 
that
RGI SSB scalar masses in Gauge-Yukawa unified models satisfy a
universal sum rule.

Consider the superpotential given by
\be
W= \frac{1}{2}\,\mu^{ij} \,\Phi_{i}\,\Phi_{j}+
\frac{1}{6}\,C^{ijk} \,\Phi_{i}\,\Phi_{j}\,\Phi_{k}~,
\label{supot}
\ee
where $\mu^{ij}$ (the mass terms) and $C^{ijk}$ (the Yukawa couplings) are
gauge invariant tensors and 
the matter field $\Phi_{i}$ transforms
according to the irreducible representation  $R_{i}$
of the gauge group $G$.  The Lagrangian for SSB terms is 
\be
\label{SSB-terms}
-{\cal L}_{\rm SSB} =
\frac{1}{6} \,h^{ijk}\,\phi_i \phi_j \phi_k
+
\frac{1}{2} \,b^{ij}\,\phi_i \phi_j
+
\frac{1}{2} \,(m^2)^{j}_{i}\,\phi^{*\,i} \phi_j+
\frac{1}{2} \,M\,\lambda \lambda+\mbox{H.c.},
\ee
where the $\phi_i$ are the scalar parts of the chiral superfields
$\Phi_i, ~\lambda$ are the gauginos and $M$ their unified mass,
$h^{ijk}$ and $b^{ij}$ are the trilinear and bilinear dimensionful
couplings respectively, and $(m^2)^{j}_{i}$ the soft scalars masses.

Let us recall that the one-loop $\beta$-function of the gauge coupling
$g$ is given by
\cite{Parkes:1984dh,West:1984dg,Jones:1985ay,Jones:1984cx,Parkes:1985hh}
  \bea \beta^{(1)}_{g}=\frac{d g}{d t} =
  \frac{g^3}{16\pi^2}\,[\,\sum_{i}\,T(R_{i})-3\,C_{2}(G)\,]~,
\label{betag}
\eea
where $C_{2}(G)$ is the quadratic Casimir of the adjoint representation of the associated
gauge group $G$. $T(R)$ is given by the relation $\textrm{Tr}[T^aT^b]=T(R)\delta^{ab}$, where
$T^a$ are the generators of the group in the appropriate representation.
Similarly the $\beta$-functions of $C_{ijk}$, by virtue of the non-renormalization theorem, are related to the
anomalous dimension matrix $\gamma^i_j$ of the chiral superfields as:
\be
\beta_C^{ijk} =
  \frac{d C_{ijk}}{d t}~=~C_{ijl}\,\gamma^{l}_{k}+
  C_{ikl}\,\gamma^{l}_{j}+
  C_{jkl}\,\gamma^{l}_{i}~.
\label{betay}
\ee At one-loop level the anomalous dimension, $\gamma^{(1)}\,^i_j$ of
the chiral superfield is \cite{Parkes:1984dh,West:1984dg,Jones:1985ay,Jones:1984cx,Parkes:1985hh}
\be \gamma^{(1)}\,^i_j=\frac{1}{32\pi^2}\,[\,
C^{ikl}\,C_{jkl}-2\,g^2\,C_{2}(R)\delta^i_j\,],
\label{gamay}
\ee
where $C_{2}(R)$ is the quadratic Casimir of the
representation $R_i$, and $C^{ijk}=C_{ijk}^{*}$.
Then, the $N = 1$ non-renormalization theorem
\cite{Wess:1973kz,Iliopoulos:1974zv,
Fujikawa:1974ay} ensures
there are no extra mass and cubic-interaction-term renormalizations,
implying that the $\beta$-functions of $C_{ijk}$ can be expressed as
linear combinations of the anomalous dimensions $\gamma^i_j$.

Here we assume that the reduction equations admit power series solutions of the form
\be
C^{ijk} = g\,\sum_{n=0}\,\rho^{ijk}_{(n)} g^{2n}~.
\label{Yg}
\ee

In order to obtain higher-loop results instead of knowledge of
explicit $\beta$-functions, which anyway are known only up to
two-loops, relations among $\beta$-functions are required.

The progress made using the spurion technique,
\cite{Delbourgo:1974jg,Salam:1974pp,Grisaru:1979wc,
Fujikawa:1974ay,Girardello:1981wz} leads to
all-loop relations among SSB $\beta$-functions 
\cite{Hisano:1997ua,Jack:1997pa,Avdeev:1997vx,Kazakov:1998uj,
Kazakov:1997nf,Jack:1997eh}. 
The assumption, following \cite{Jack:1997pa}, that the relation among couplings
\be
h^{ijk} = -M (C^{ijk})'
\equiv -M \frac{d C^{ijk}(g)}{d \ln g}~,
\label{h2}
\ee
is RGI and furthermore, the use the all-loop gauge $\beta$-function of
Novikov {\em et al.}\cite{Novikov:1983ee,Novikov:1985rd}
\be
\beta_g^{\rm NSVZ} =
\frac{g^3}{16\pi^2}
\left[ \frac{\sum_l T(R_l)(1-\gamma_l /2)
-3 C_2(G)}{ 1-g^2C_2(G)/8\pi^2}\right]~,
\label{bnsvz}
\ee
lead to the all-loop RGI sum rule \cite{Kobayashi:1998jq} (assuming $(m^2)^i{}_j=m^2_j\delta^i_j$),
\begin{equation}
\begin{split}
m^2_i+m^2_j+m^2_k &=
|M|^2 \left\{~
\frac{1}{1-g^2 C_2(G)/(8\pi^2)}\frac{d \ln C^{ijk}}{d \ln g}
+\frac{1}{2}\frac{d^2 \ln C^{ijk}}{d (\ln g)^2}~\right\}\\
& +\sum_l
\frac{m^2_l T(R_l)}{C_2(G)-8\pi^2/g^2}
\frac{d \ln C^{ijk}}{d \ln g}~.
\label{sum2}
\end{split}
\end{equation}

The all-loop results on the SSB $\beta$-functions 
lead to all-loop RGI relations (see e.g. \cite{Mondragon:2013aea}). If we assume:\\
(a) the existence of a RGI surfaces on which $C = C(g)$, or equivalently that
\be
\label{Cbeta}
\frac{dC^{ijk}}{dg} = \frac{\beta^{ijk}_C}{\beta_g}
\ee
holds,  i.e. reduction of couplings is possible, and\\
(b) the existence of a RGI surface on which
\be
\label{h2NEW}
h^{ijk} = - M \frac{dC(g)^{ijk}}{d\ln g}
\ee
holds too in all-orders, then one can prove  that the following relations are RGI to all-loops \cite{Jack:1999aj,Kobayashi:1998iaa} (note that in the above assumptions (a) and (b) 
 we do not rely on specific solutions to these equations)
\begin{align}
M &= M_0~\frac{\beta_g}{g} ,  \label{Mbeta} \\
h^{ijk}&=-M_0~\beta_C^{ijk},  \label{hbeta}  \\
b^{ij}&=-M_0~\beta_{\mu}^{ij},\label{bij}\\
(m^2)^i{}_j&= \frac{1}{2}~|M_0|^2~\mu\frac{d\gamma^i{}_j}{d\mu},
\label{scalmass}
\end{align}
where $M_0$ is an arbitrary reference mass scale to be specified
shortly. 

Finally we would like to emphasize that under the same assumptions (a)
and (b) the sum rule given in Eq.(\ref{sum2}) has been proven
\cite{Kobayashi:1998jq} to be all-loop RGI, which  gives us
a generalization of Eq.(\ref{scalmass}) to be applied in
considerations of non-universal soft scalar masses, which are
necessary in many cases including the MSSM.

As it was emphasized in ref \cite{Jack:1999aj} the set of the all-loop
RGI relations (\ref{Mbeta})-(\ref{scalmass}) is the one obtained in
the \textit{Anomaly Mediated SB Scenario}
\cite{Randall:1998uk,Giudice:1998xp}, 
by fixing
the $M_0$ to be $m_{3/2}$, which is the natural scale in the
supergravity framework.
A final remark concerns the resolution of the fatal problem of the
anomaly induced scenario in the supergravity framework, which  is
here solved thanks to the sum rule (\ref{sum2}). Other solutions have been provided by introducing
Fayet-Iliopoulos terms \cite{Hodgson:2005en}.

\section{Applications of the Reduction of Couplings Method}

In this section we show how to apply the reduction of couplings method
in two scenarios, the MSSM and Finite Unified Theories. We will apply
it only to the third generation of fermions and in the soft
supersymmetry breaking terms. After the reduction of couplings takes
place, we are left with relations at the unification scale for the
Yukawa couplings of the quarks in terms of the gauge coupling
according to Eq.~(\ref{Yg}), for the trlininear terms in terms of the
Yukawa couplings and the unified gaugino mass Eq.~(\ref{h2NEW}), and a
sum rule for the soft scalar masses also proportional to the unified
gaugino mass Eq.~(\ref{sum2}), as applied in each model.

\subsection{RE in the MSSM}

We will examine here the reduction of couplings method  applied to
the MSSM, which  is defined by the superpotential,
\begin{equation}
\label{supot2}
W = Y_tH_2Qt^c+Y_bH_1Qb^c+Y_\tau H_1L\tau^c+ \mu H_1H_2    ,    
\end{equation}
with soft breaking terms,
\begin{equation}
\label{}
\begin{split}
-\mathcal{L}_{SSB} &= \sum_\phi m^2_\phi\phi^*\phi+
\left[m^2_3H_1H_2+\sum_{i=1}^3 \frac 12 M_i\lambda_i\lambda_i +\textrm{h.c}\right]\\
&+\left[h_tH_2Qt^c+h_bH_1Qb^c+h_\tau H_1L\tau^c+\textrm{h.c.}\right] ,          
\end{split}
\end{equation}
where the last line refers to the scalar components of the corresponding superfield.
In general $Y_{t,b,\tau}$ and $h_{t,b,\tau}$ are $3\times 3$ matrices, but we work throughout in the
approximation that the  matrices are diagonal, and neglect the couplings
of the first two generations.

Assuming perturbative expansion of all three Yukawa couplings in
favour of $g_3$ satisfying the reduction equations we find that the
coefficients of the $Y_\tau$ coupling turn imaginary. Therefore, we
take $Y_\tau$ at the GUT scale to be an independent variable. Thus, in
the application of the reduction of couplings in the MSSM that we
examine here, in the first stage we neglect the Yukawa couplings of
the first two generations, while we keep $Y_\tau$ and the gauge
couplings $g_2$ and $g_1$, which cannot be reduced consistently, as
corrections. This ``reduced'' system holds
at all scales, and thus serve as boundary conditions of the RGEs of
the MSSM at the unification scale, where we assume that the gauge
couplings meet \cite{Mondragon:2013aea}.

In that case, the coefficients of the expansions
(again at the unification scale) 
\be
\frac{Y_t^2}{4\pi}=c_1\frac{g_3^2}{4\pi}+c_2\left(\frac{g_3^2}{4\pi}\right)^2\label{alpha_t}~; \quad
\frac{Y_b^2}{4\pi}=p_1\frac{g_3^2}{4\pi}+p_2\left(\frac{g_3^2}{4\pi}\right)^2 
\ee
are given by
\be
\label{app_atau}
\begin{split}
c_1&=\frac{157}{175}+\frac 1{35}K_\tau=0.897 +0.029K_\tau
\\
p_1&=\frac{143}{175}-\frac{6}{35}K_\tau=0.817 -0.171K_\tau
\\
c_2&=\frac 1{4\pi}\,
\frac{1457.55 - 84.491 K_\tau - 9.66181 K_\tau^2 - 0.174927 K_\tau^3}
{818.943 - 89.2143 K_\tau - 2.14286 K_\tau^2}\\
p_2&=\frac 1{4\pi}\,
\frac{1402.52 - 223.777 K_\tau - 13.9475 K_\tau^2 - 0.174927 K_\tau^3}
{818.943 - 89.2143 K_\tau - 2.14286 K_\tau^2}
\end{split}
\ee
where
\be
K_\tau=Y_\tau^2/g_3^2
\label{ktau}
\ee
The couplings $Y_t$,$Y_b$ and
$g_3$ are not only reduced, but  they provide predictions consistent
with the observed experimental values, as we will show in subsection 4.2. According to the analysis
presented in Section 2 
the RGI relations in the SSB sector hold, assuming the existence of RGI
surfaces where Eqs.(\ref{Cbeta}) and (\ref{h2NEW}) are valid. 

Since all gauge couplings in the MSSM meet at the unification
point, we are led to the following boundary conditions at the
unification scale:
\begin{align}
Y_t^2&=c_1 g_U^2 + c_2 g_U^4/(4\pi) \quad\textrm{and}\quad
Y_b^2=p_1 g_U^2 + p_2 g_U^4/(4\pi)
\label{Y_t_Y_b} \\
h_{t,b} &= - M_U Y_{t,b}   , \label{eq:htb-MU}
\\
m_3^2 &= - M_U \mu          ,\label{m32}
\end{align}
where  $M_U$ is the unification scale, $c_{1,2}$ and $p_{1,2}$ are the solutions of the algebraic
system of the two reduction equations  taken at the unification 
scale (while keeping only the first term\footnote{The second term can
  be determined once the first term is known.} of the perturbative
expansion of the Yukawas in favour of $g_3$ for
Eqs.(\ref{eq:htb-MU}) and (\ref{m32})), and a set of
equations resulting from the application of the sum rule
\be
m_{H_2}^2 + m_Q^2 + m_{t^c}^2 = M_U^2,\label{sum_rule_1} \quad
m_{H_1}^2 + m_Q^2 + m_{b^c}^2 = M_U^2 ,
\ee
noting that the sum rule introduces four free parameters.

\subsection{Predictions of the Reduced MSSM}
 With these boundary conditions we run the MSSM RGEs
down to the SUSY scale, which we take to be the geometrical average of
the stop masses, and then run the SM RGEs down to the electroweak
scale ($M_Z$), where we compare with the experimental values of the
third generation quark masses.  The RGEs are taken at two-loops for
the gauge and Yukawa couplings and at one-loop for the soft breaking
parameters.  We let $M_U$ and $|\mu|$ at the unification scale to vary
between $\sim 1 \tev \sim 11 \tev$, for the two possible signs of
$\mu$. In evaluating the $\tau$ and bottom masses we have taken into
account the one-loop radiative corrections that come from the SUSY
breaking \cite{Carena:1994bv,Guasch:2001wv}; 
 in particular for large $\tan\beta$ they can give 
sizeable contributions to the bottom quark mass.

Recall that
$Y_{\tau}$ is not reduced  and is a free parameter in this analysis.
The parameter $K_\tau$, related to  $Y_{\tau}$  through Eq.~(\ref{ktau}) is further constrained by allowing only the
values that are also compatible  with the top and bottom quark masses
simultaneously 
within 1 and $2\sigma$ of their central experimental value.
In  the case that  $\textrm{sign}(\mu)=+$,  there is no value for $K_\tau$ where both the top and the bottom
quark masses agree simultaneously with their experimental value,
therefore we only consider the negative sign of $\mu$ from now on.
We use the experimental value of the top quark pole mass as
\be
m_t^{\rm exp} = (173.2 \pm 0.9) \gev ~.
\label{topmass-exp}
\ee
The bottom mass is calculated at $M_Z$ to avoid uncertainties that
come from running down to the pole mass and, as previously mentioned,
the SUSY radiative corrections both to the tau and the bottom quark masses
have been taken into account \cite{Nakamura:2010zzi}
\be 
m_b(M_Z) =
(2.83 \pm 0.10) \gev . 
\label{botmass-MZ}
 \ee
The variation of $K_\tau$ is in
the range $\sim 0.33 \sim 0.5$ if the  agreement with both top and bottom masses is at the 2$\sigma$ level.

\begin{figure}[t]
\begin{center}
\includegraphics[scale=0.3]{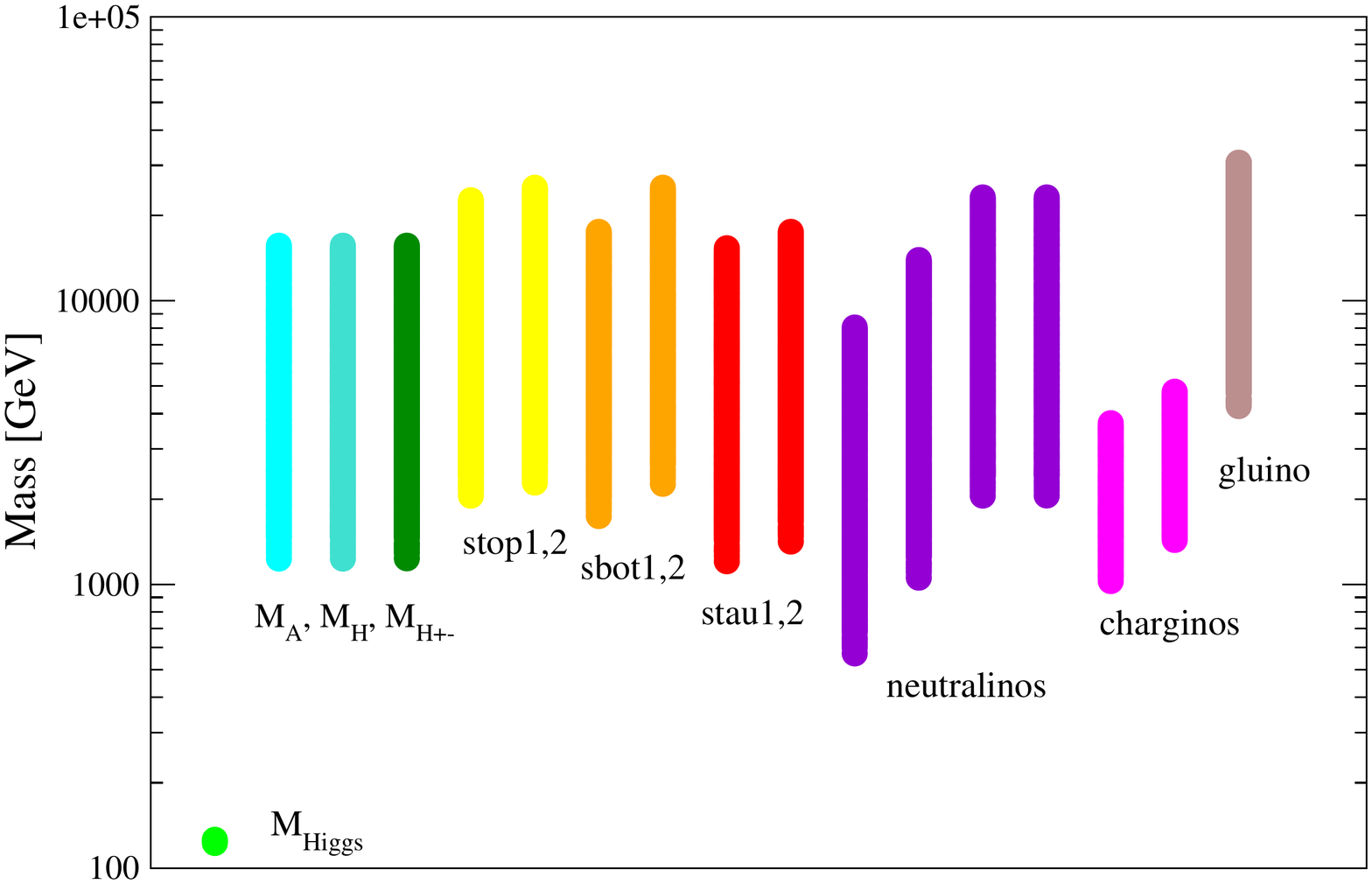}\hspace{0.5cm}\includegraphics[scale=0.3]{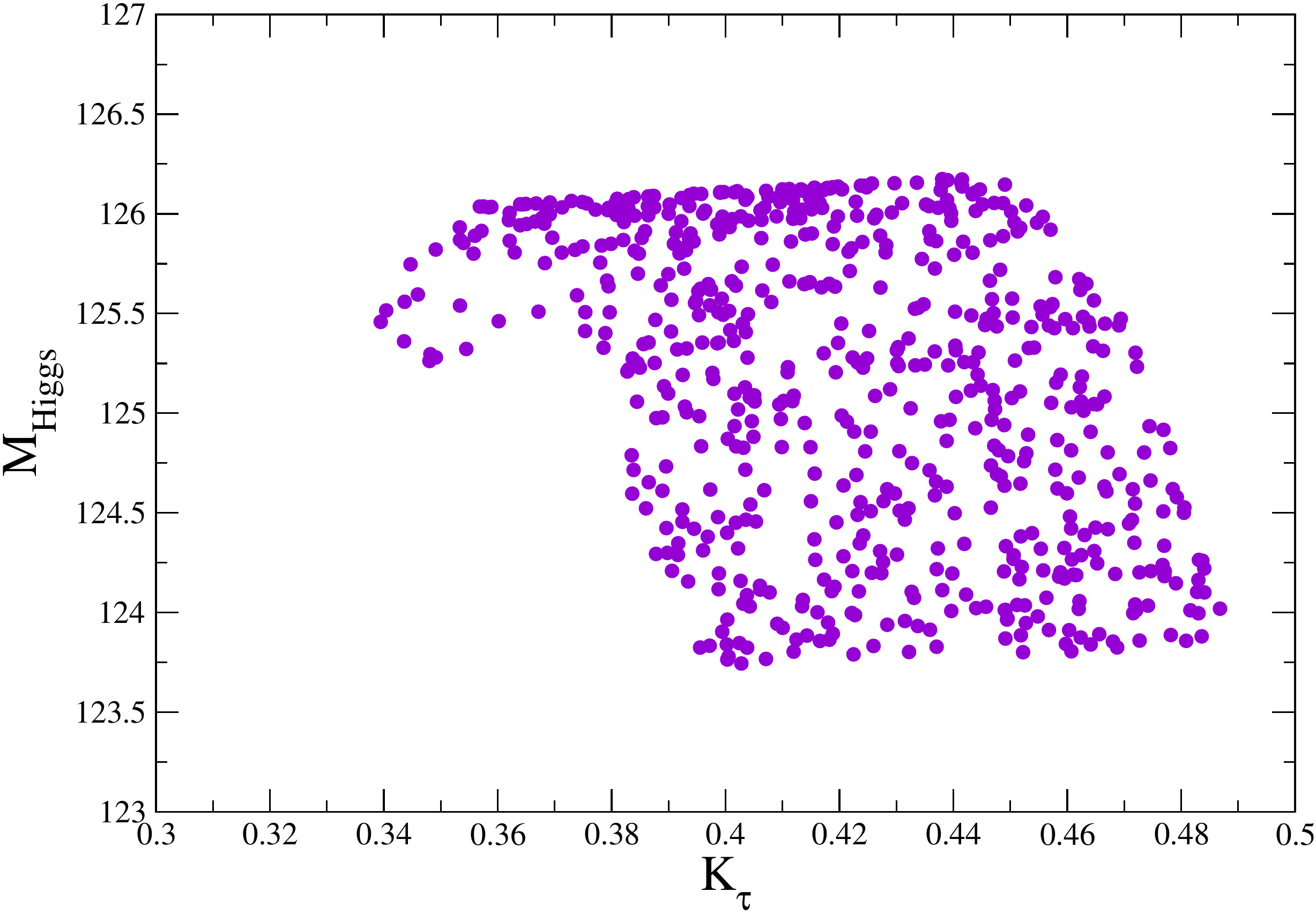}
\caption{The left plot shows the SUSY spectrum as function of the
  reduced MSSM. From left to right are shown: The lightest Higgs, the
  pseudoscalar one $M_A$, the heavy neutral one $M_H$, the two charged 
  Higgses $M_{H^\pm}$; then come the two stops,  two sbottoms and
  two staus, the four neutralinos, the two charginos,  and at the end
  the gluino.
  The right plot shows the lightest Higgs boson mass as a function of $K_\tau=Y_\tau^2/g_3^2$~.}
\label{constrain-AKTAUvsMHiggs}
\end{center}
\end{figure}

Finally, assuming the validity of Eq.(\ref{h2NEW}) for the corresponding couplings
to those that have been reduced before, we calculate
the Higgs mass as well as the whole Higgs and sparticle spectrum using
Eqs.(\ref{Y_t_Y_b})-(\ref{sum_rule_1}),
and we 
present them in Fig.~(\ref{constrain-AKTAUvsMHiggs}) 
The Higgs mass was calculated using a ``mixed-scale'' one-loop RG
approach, which is known to approximate the leading two-loop
corrections as evaluated by the full diagrammatic calculation
\cite{Carena:2000dp,Heinemeyer:2004ms}. 
However, more refined Higgs mass calculations, and in particular the
results of \cite{Hahn:2013ria} are not (yet) included.


In the left plot of Fig.(\ref{constrain-AKTAUvsMHiggs}) we show the
full mass spectrum of the model. We find that the masses of the heavier
Higgses have relatively high values, above the TeV scale. In addition
we find a generally heavy supersymmetric spectrum starting with a
neutralino as LSP at $\sim 500$ GeV and comfortable agreement with the
LHC bounds due to the non-observation of  coloured supersymmetric
particles~\cite{Chatrchyan:2012vp,Pravalorio:susy2012,Campagnari:susy2012}. 
Finally note that although the $\mu < 0$ found in our analysis would disfavour
the model in connection with the anomalous magnetic moment of the muon, such
a heavy spectrum gives only a negligible correction to the SM prediction.
We plan to extend our analysis by examining the restrictions that will be
imposed in the spectrum by the $B$-physics and Cold Dark Matter (CDM) constraints.

In the right plot of Fig.~(\ref{constrain-AKTAUvsMHiggs}) we show the
results for the light Higgs boson mass as a function of $K_\tau$. The
results are in the range
$123.7$ - $126.3$~GeV, where the 
uncertainty is due to the variation of $K_\tau$, the gaugino mass
$M_U$ and the variation of the scalar soft masses, which are however
constrained by the sum rules (\ref{sum_rule_1}).  The gaugino mass $M_U$ is in the range $\sim
1.3\tev \sim 11 \tev$, the lower values having been discarded since
they do not allow for radiative electroweak symmetry breaking. To the lightest Higgs mass value one has to add at least $\pm 2$~GeV
coming from unknown higher order corrections \cite{Degrassi:2002fi}.
Therefore it is in excellent agreement with the experimental results
of ATLAS and CMS \cite{Aad:2012tfa,ATLAS:2013mma,Chatrchyan:2012ufa,Chatrchyan:2013lba}. 

\subsection{Finiteness}
Finiteness can be understood by considering a chiral, anomaly free,
$N=1$ globally supersymmetric 
gauge theory based on a group $G$ with gauge coupling
constant $g$.  Consider the superpotential Eq.~(\ref{supot}) together
with the soft supersymmetry breaking Lagrangian Eq.~(\ref{SSB-terms}).
All the one-loop $\beta$-functions of the theory
vanish if  the $\beta$-function of the gauge coupling $\beta_g^{(1)}$, and
the anomalous dimensions of the Yukawa couplings $\gamma_i^{j(1)}$, vanish, i.e.
\begin{equation}
\sum _i \ell (R_i) = 3 C_2(G) \,,~
\frac{1}{2}C_{ipq} C^{jpq} = 2\delta _i^j g^2  C_2(R)\ ,
\label{2}
\end{equation}
where $\ell (R_i)$ is the Dynkin index of $R_i$,  and $C_2(G)$ and
$C_{2}(R)$  are the
quadratic Casimir invariants of the adjoint representation of $G$ and
the representation $R_i$, respectively.
These conditions are also enough to guarantee two-loop finiteness 
\cite{Jones:1984cu}.  A striking fact is the existence of
a theorem \cite{Lucchesi:1987he,Piguet:1986td,Lucchesi:1996ir}, 
that 
guarantees the vanishing of the $\beta$-functions to all-orders in
perturbation theory.  This requires that, in addition to the one-loop
finiteness conditions (\ref{2}), the Yukawa couplings are reduced in
favour of the gauge coupling to all-orders (see \cite{Heinemeyer:2011kd}
for details).  Alternatively, similar results can be obtained
\cite{Ermushev:1986cu,Kazakov:1987vg,Leigh:1995ep} 
using an analysis of the all-loop NSVZ gauge beta-function
\cite{Novikov:1983ee,Shifman:1996iy}.

Since we would like to consider 
only finite theories here, we assume that 
the gauge group is  a simple group and the one-loop
$\beta$-function of the 
gauge coupling $g$  vanishes.\footnote{
Realistic Finite Unified Theories based on product
gauge groups, where the finiteness implies three generations of matter, have
also been studied\cite{Ma:2004mi,Heinemeyer:2010zza}.}
We also assume that the reduction equations 
admit power series solutions of the form Eq.~(\ref{Yg}).
According to the finiteness theorem
of ref.~\cite{Lucchesi:1987ef,Lucchesi:1987he,Piguet:1986td,Lucchesi:1996ir}, the theory is then finite to all orders in
perturbation theory, if, among others, the one-loop anomalous dimensions
$\gamma_{i}^{j(1)}$ vanish.  The one- and two-loop finiteness for
$h^{ijk}$ can be achieved through the relation \cite{Jack:1994kd}
\bea h^{ijk} &=& -M C^{ijk}+\dots =-M
\rho^{ijk}_{(0)}\,g+O(g^5)~, 
\label{hY}
\eea
where $\dots$ stand for  higher order terms.

In addition it was found that the  RGI SSB scalar masses
in Gauge-Yukawa unified models satisfy a universal sum rule at
one-loop \cite{ Kawamura:1997cw}. This result was generalized to two-loops
for finite theories \cite{Kobayashi:1997qx}, and then to all-loops for
general Gauge-Yukawa and finite unified theories \cite{Kobayashi:1998jq}.
From these latter results,  the following soft scalar-mass sum rule is found
\cite{Kobayashi:1997qx}
\begin{equation}
\frac{(~m_{i}^{2}+m_{j}^{2}+m_{k}^{2}~)}{M M^{\dag}} =
1+\frac{g^2}{16 \pi^2}\,\Delta^{(2)}
+O(g^4)~
\label{zoup-sumr}
\end{equation}
for i, j, k with $\rho^{ijk}_{(0)} \neq 0$, where  $m_{i,j,k}^{2}$ are
the scalar masses and $\Delta^{(2)}$ is
the two-loop correction 
which vanishes for the
universal choice, i.e.\ when all the soft scalar masses are the same at
the unification point, as well as in the model 
considered here.

\subsection{\boldmath An {$SU(5)$} Finite Unified Theory}

We examine an  all-loop Finite Unified theory 
with $SU(5)$ as gauge group, where the reduction of couplings has been
applied to the third generation of quarks and leptons. 
The particle content of the model we will study, which we denote \FUTB\, consists of the
following supermultiplets: three ($\overline{\bf 5} + \bf{10}$),
needed for each of the three generations of quarks and leptons, four
($\overline{\bf 5} + {\bf 5}$) and one ${\bf 24}$ considered as Higgs
supermultiplets. 
When the gauge group of the finite GUT is broken the theory is no
longer finite, and we will assume that we are left with the MSSM
\cite{Kapetanakis:1992vx,Kubo:1994bj,Kubo:1994xa,Kubo:1995hm,Kubo:1997fi}.

A predictive Gauge-Yukawa unified $SU(5)$ model which is finite to all
orders, in addition to the requirements mentioned already, should also
have the following properties:
\begin{enumerate}
\item 
One-loop anomalous dimensions are diagonal,
i.e.,  $\gamma_{i}^{(1)\,j} \propto \delta^{j}_{i} $. 
\item Three fermion generations, in the irreducible representations
  $\overline{\bf 5}_{i},{\bf 10}_i~(i=1,2,3)$, which obviously should
  not couple to the adjoint ${\bf 24}$.
\item The two Higgs doublets of the MSSM should mostly be made out of a
pair of Higgs quintet and anti-quintet, which couple to the third
generation.
\end{enumerate}

After the reduction
of couplings the symmetry is enhanced, leading to the 
following superpotential \cite{Mondragon:2009zz}
\bea
W &=& \sum_{i=1}^{3}\,[~\frac{1}{2}g_{i}^{u}
\,{\bf 10}_i{\bf 10}_i H_{i}+
g_{i}^{d}\,{\bf 10}_i \overline{\bf 5}_{i}\,
\overline{H}_{i}~] +
g_{23}^{u}\,{\bf 10}_2{\bf 10}_3 H_{4} \\
 & &+g_{23}^{d}\,{\bf 10}_2 \overline{\bf 5}_{3}\,
\overline{H}_{4}+
g_{32}^{d}\,{\bf 10}_3 \overline{\bf 5}_{2}\,
\overline{H}_{4}+
g_{2}^{f}\,H_{2}\, 
{\bf 24}\,\overline{H}_{2}+ g_{3}^{f}\,H_{3}\, 
{\bf 24}\,\overline{H}_{3}+
\frac{g^{\lambda}}{3}\,({\bf 24})^3~.\nonumber
\label{w-futb}
\eea
The non-degenerate and isolated solutions to
$\gamma^{(1)}_{i}=0$ give us: 
\bea 
&& (g_{1}^{u})^2
=\frac{8}{5}~ g^2~, ~(g_{1}^{d})^2
=\frac{6}{5}~g^2~,~
(g_{2}^{u})^2=(g_{3}^{u})^2=\frac{4}{5}~g^2~,\label{zoup-SOL52}\\
&& (g_{2}^{d})^2 = (g_{3}^{d})^2=\frac{3}{5}~g^2~,~
(g_{23}^{u})^2 =\frac{4}{5}~g^2~,~
(g_{23}^{d})^2=(g_{32}^{d})^2=\frac{3}{5}~g^2~,
\nonumber\\
&& (g^{\lambda})^2 =\frac{15}{7}g^2~,~ (g_{2}^{f})^2
=(g_{3}^{f})^2=\frac{1}{2}~g^2~,~ (g_{1}^{f})^2=0~,~
(g_{4}^{f})^2=0~,\nonumber 
\eea 
and from the sum rule we obtain:
\be
m^{2}_{H_u}+
2  m^{2}_{{\bf 10}} =M^2~,~
m^{2}_{H_d}-2m^{2}_{{\bf 10}}=-\frac{M^2}{3}~,~
m^{2}_{\overline{{\bf 5}}}+
3m^{2}_{{\bf 10}}=\frac{4M^2}{3}~,
\label{sumrB}
\ee
i.e., in this case we have only two free parameters  
$m_{{\bf 10}}$  and $M$ for the dimensionful sector.

As already mentioned, after the $SU(5)$ gauge symmetry breaking we
assume we have the MSSM, i.e. only two Higgs doublets.  This can be
achieved by introducing appropriate mass terms that allow to perform a
rotation of the Higgs sector
\cite{Leon:1985jm,Kapetanakis:1992vx,Kubo:1994bj,Kubo:1994xa,Kubo:1995hm,Mondragon:1993tw,Hamidi:1984gd, Jones:1984qd},
in such a way that only one pair of Higgs doublets, coupled mostly to
the third family, remains light and acquire vacuum expectation values.
To avoid fast proton decay the usual fine tuning to achieve
doublet-triplet splitting is performed, although the mechanism is not
identical to minimal $SU(5)$, since we have an extended Higgs sector.

Thus, after the gauge symmetry of the GUT theory is broken we are left
with the MSSM, with the boundary conditions for the third family given
by the finiteness conditions, while the other two families are not
restricted.

\subsection{Predictions of the Finite Model}
 
Since the gauge symmetry is spontaneously broken below $M_{\rm GUT}$,
the finiteness conditions do not restrict the renormalization
properties at low energies, and all it remains are boundary conditions
on the gauge and Yukawa couplings 
(\ref{zoup-SOL52}), the $h=-MC$ (\ref{hY}) relation, and the soft
scalar-mass sum rule at $M_{\rm GUT}$.  The analysis follows along the
same lines as in the MSSM case.

\begin{figure}
\begin{center}
          \includegraphics[scale=0.3]{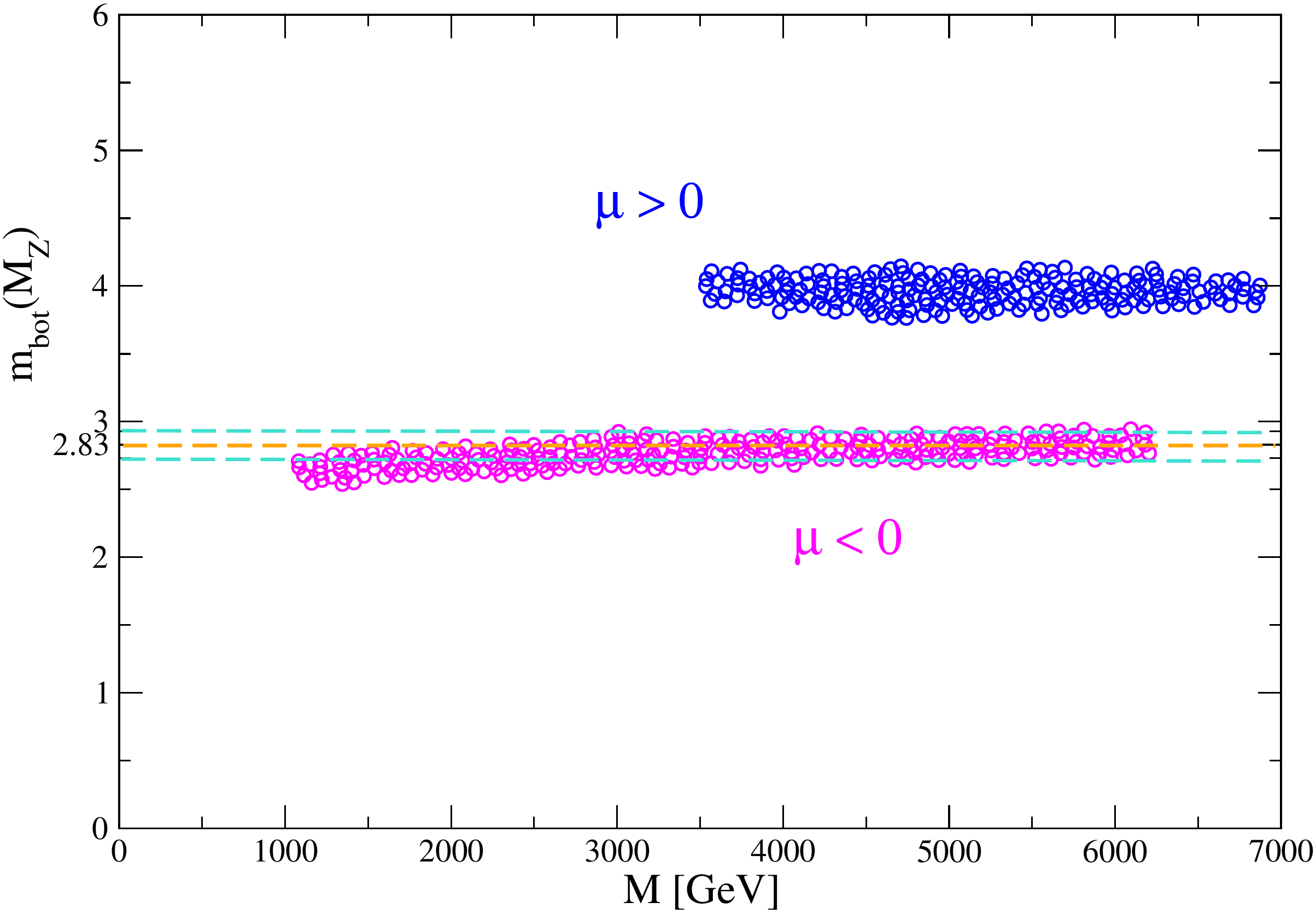}\hspace{0.5cm}
          \includegraphics[scale=0.3]{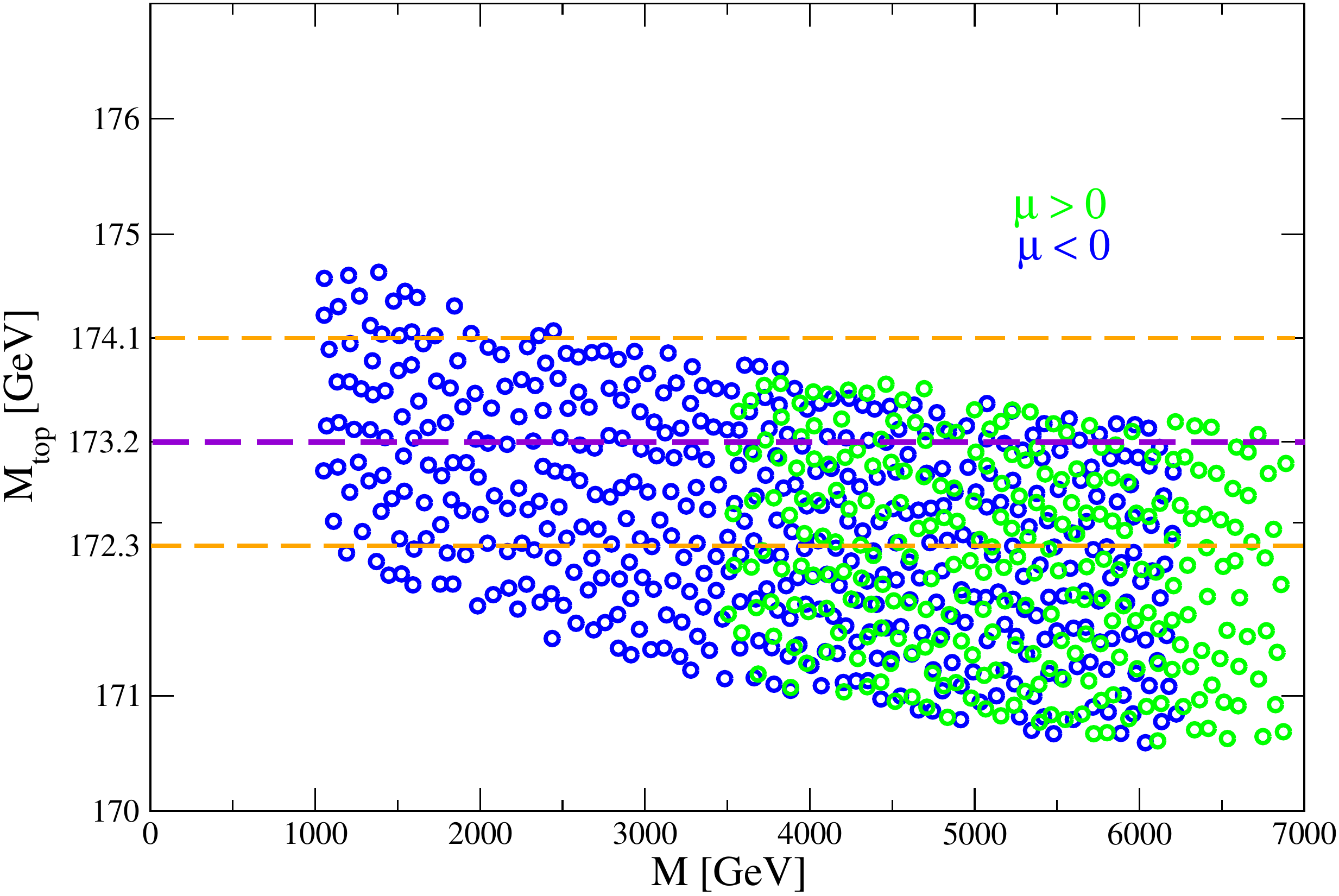}
       \caption{The bottom quark mass at the $Z$~boson scale (left) 
                and top quark pole mass (right) are shown 
                as function of $M$, the unified gaugino mass.}
\label{fig:MtopbotvsM}
\end{center}
\end{figure}

In Fig.\ref{fig:MtopbotvsM} we show the {\bf FUT}
predictions for $\mt$ and $\mb (M_Z)$ as a function of the unified 
gaugino mass $M$, for the two cases $\mu <0$ and $\mu >0$. 
The bounds on the $\mb(M_Z)$ and the $\mt$ mass clearly single out
$\mu <0$, as the solution most compatible with these 
experimental constraints \cite{Heinemeyer:2007tz,Heinemeyer:2010xt}.  

We now analyze the impact of further low-energy observables on the model
{\bf FUT} with $\mu < 0$.
As  additional constraints we consider 
the flavour observables $\brat(b \to s \gamma)$ and $\brat(B_s \to \mu^+ \mu^-)$.

For the branching ratio $\brat(b \to s \gamma)$, we take the value
given by the Heavy Flavour Averaging Group (HFAG) is~\cite{bsgexp}
\beq 
\brat(b \to s \gamma ) = (3.55 \pm 0.24 {}^{+0.09}_{-0.10} \pm
0.03) \times 10^{-4} .
\label{bsgaexp}
\eeq 
For the branching ratio $\brat(B_s \to \mu^+ \mu^-)$, the SM 
prediction is at the level of $10^{-9}$, while we employ an upper
limit of 
\beq 
\brat(B_s \to \mu^+ \mu^-) \lsim 4.5 \times 10^{-9} 
\eeq 
at the $95\%$ C.L.~\cite{Aaij:2012ac}.
This is in relatively good agreement with the recent direct measurement of
this quantity by CMS and LHCb~\cite{CMSLHCb}. 
As we do not expect a sizable impact of the new measurement on our results, we
stick for our analysis to the simple upper limit.

\begin{figure}[t]
\begin{center}
           \includegraphics[scale=0.3]{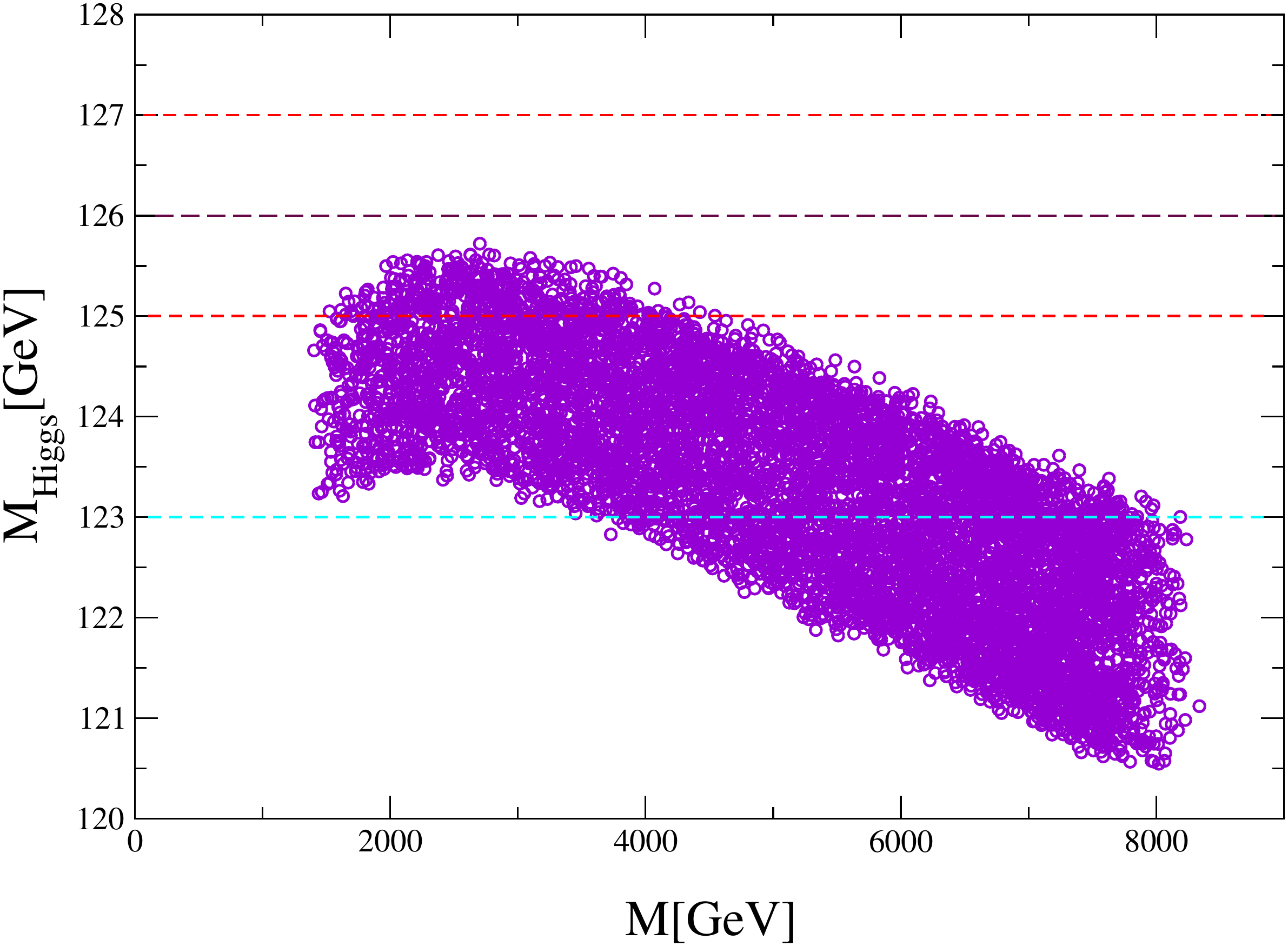}
        \caption{The lightest Higgs mass, $M_h$,  as function of $M$ for
          the model {\bf FUT} with $\mu < 0$.}
\label{fig:Higgs}
\end{center}
\end{figure}

For the lightest Higgs mass prediction we used the code {\tt
  FeynHiggs}~\cite{Heinemeyer:1998yj,Heinemeyer:1998np,Frank:2006yh,
Degrassi:2002fi}.
The prediction for $M_h$ of {\bf FUT} with $\mu < 0$ is shown in
Fig.~\ref{fig:Higgs}, where the constraints from the two $B$~physics
observables are taken into account.  The lightest Higgs mass ranges in
\beq
M_h \sim 121-126 \gev~ , 
\label{eq:Mhpred}
\eeq
where the uncertainty comes from variations of the soft scalar masses.
To this value one has to add at least $\pm 2$ GeV coming from unkonwn
higher order corrections~\cite{Degrassi:2002fi}\footnote{We have not
  yet taken into account the improved $\Mh$ prediction presented
  in~\cite{Hahn:2013ria} (and implemented into the most recent version
  of {\tt FeynHiggs}), which will lead to an upward shift in the Higgs
  boson mass prediction.}.  We have also included a small variation,
due to threshold corrections at the GUT scale, of up to $5 \%$ of the
FUT boundary conditions.  The masses of the heavier Higgs bosons are
found at higher values in comparison with our previous
analyses~\cite{Heinemeyer:2007tz,Heinemeyer:2008qw,Heinemeyer:2009zz,Heinemeyer:2012sy}.
This is due to the more stringent bound on $\br(B_s \to \mu^+\mu^-)$,
which pushes the heavy Higgs masses beyond $\sim 1 \tev$, excluding
their discovery at the LHC.

We  impose now a further 
constraint on our results, which is  the value of the Higgs mass 
\beq 
M_h \sim 126.0 \pm 1 \pm 2 \gev~ ,
\label{eq:Mh125}
\eeq
where $\pm 3 \gev$ corresponds to the current theory and experimental
uncertainty, and $\pm 1 \gev$ to a reduced theory uncertainty in the
future.\footnote{In this analysis the new $\Mh$
  evaluation~\cite{Hahn:2013ria} may have a relevant impact on the
  restrictions on the allowed SUSY parameter space shown below.}  We
find that constraining the allowed values of the Higgs mass puts a
limit on the allowed values of the unified gaugino mass, as can be
seen from \reffi{fig:Higgs}.  The red lines correspond to the
anticipated future uncertainty and restrict $2 \tev \lsim M
\lsim 5 \tev$. The blue line includes the current theoretical
uncertainty. Taking this uncertainty into account no
bound on $M$ can be placed, but many parameter points can be
discarded.

\begin{figure}[t!]
\begin{center}
          \includegraphics[scale=0.4]{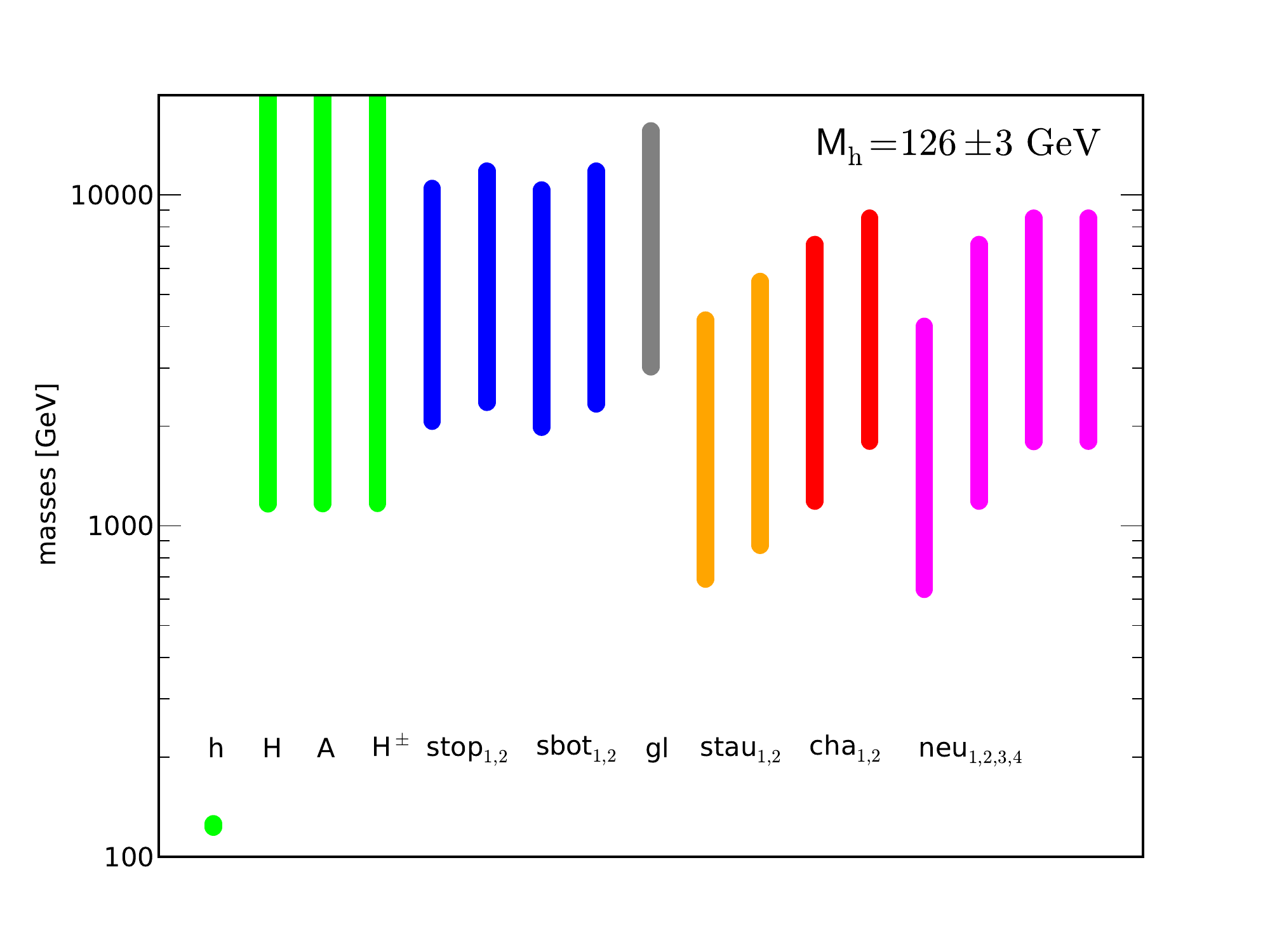}
       \includegraphics[scale=0.4]{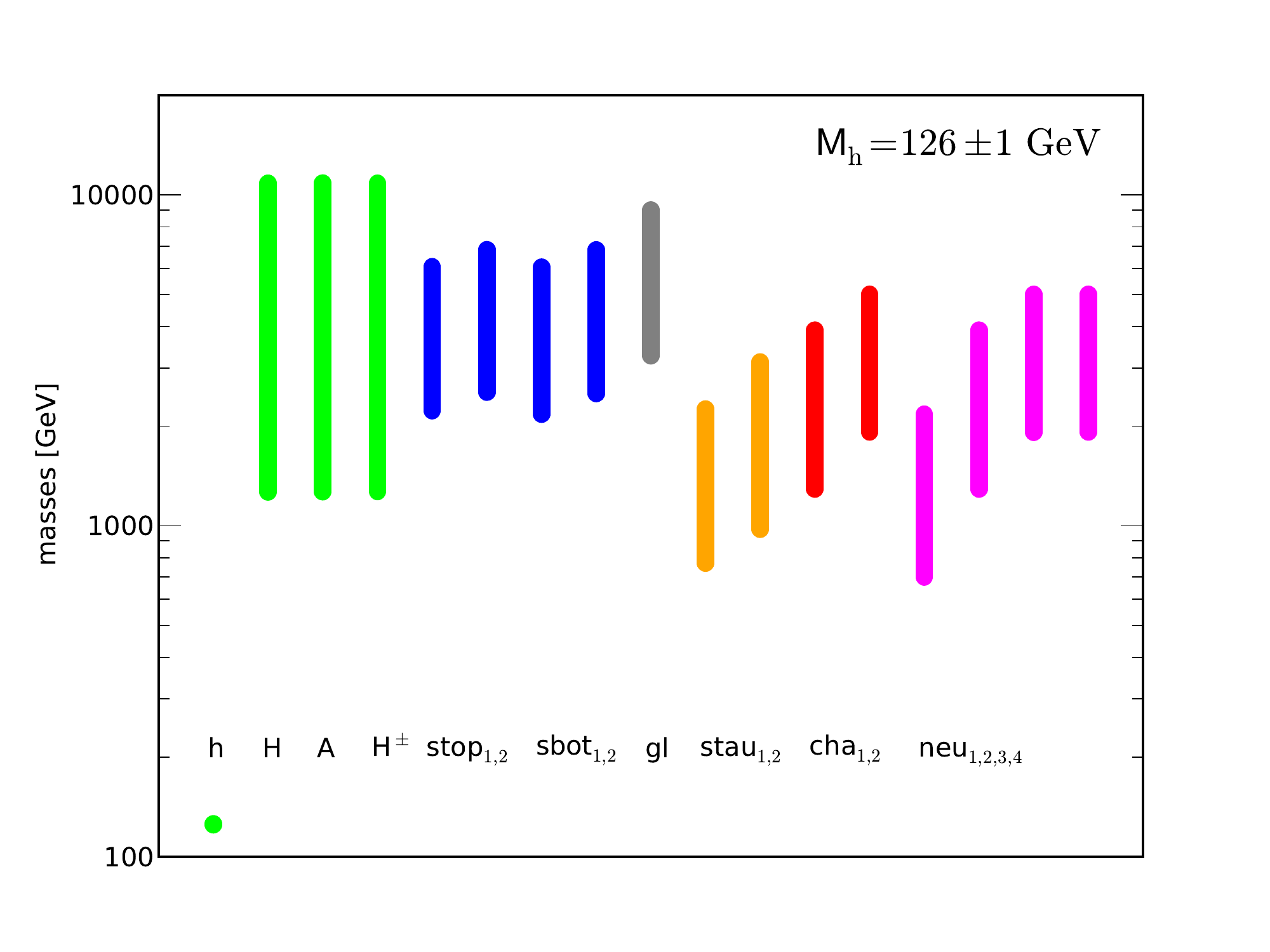}
           \caption{The left (right) plot shows the spectrum after
             imposing the constraint $\Mh = 126 \pm 3\,(1) \gev$. The
             light (green) points are the various Higgs
             boson masses, the dark (blue) points following are the
             two scalar top and bottom masses, the gray ones are 
             the gluino masses,  then come  the  scalar
             tau masses in orange (light gray), the darker (red) points to the right are the
             two chargino masses followed by the lighter shaded (pink)
             points indicating the neutralino masses.}
\label{fig:masses}
\end{center}
\end{figure}%

The full particle spectrum of model {\bf FUT} with $\mu <0$, compliant
with quark mass constraints and the $B$-physics observables is shown
in \reffi{fig:masses}. It can be seen from the figure that the
lightest observable SUSY particle (LOSP) is the light scalar tau.  In
the left (right) plot we impose $\Mh = 126 \pm 3 (1) \gev$.  Without
any $\Mh$ restrictions the coloured SUSY particles have masses above
$\sim 1.8 \tev$ in agreement with the non-observation of those
particles at the LHC
\cite{Chatrchyan:2012vp,Pravalorio:susy2012,Campagnari:susy2012}.
Including the Higgs mass constraints in general favours the lower part
of the SUSY particle mass spectra, but also cuts away the very low
values\cite{Heinemeyer:2012yj,Heinemeyer:2012ai,Heinemeyer:2013fga,Heinemeyer:2013nza}. Going
to the anticipated future theory uncertainty of $\Mh$ (as shown in the
right plot of \reffi{fig:masses}) still permits SUSY masses which would
remain unobservable at the LHC, the ILC or CLIC.  On the other hand,
large parts of the allowed spectrum of the lighter scalar tau or the
lighter neutralinos might be accessible at CLIC with $\sqrt{s} = 3
\tev$.

\section{Conclusions}
 The serious problem of the appearance of  many free parameters
in the SM of Elementary Particle Physics, takes dramatic dimensions in
the MSSM, where the free parameters are proliferated by at least
hundred more, while it is considered as the best candidate for Physics
Beyond the SM.  The idea that the Theory of Particle Physics is more
symmetric at high scales, which is broken but has remnant predictions
in the much lower scales of the SM, found its best realisation in the
framework of the MSSM assuming further a GUT beyond the scale of the
unification of couplings.  However, the unification idea, although
successful,  seems to have exhausted its potential to reduce further
the free parameters of the SM.  

A new interesting possibility towards reducing the free parameters of
a theory has been put forward in
refs. \cite{Zimmermann:1984sx,Oehme:1984yy} which consists on a
systematic search on the RGI relations among couplings. This method
might lead to further symmetry, however its scope is much wider.
After several trials it seems that the basic idea found very nice
realisations in a Finite Unified Theory and the MSSM. In the first
case one is searching for RGI relations among couplings holding beyond
the unification scale, which morever guarantee finiteness to
all-orders in perturbation theory. In the second, the search of RGI
relations among couplings is concentrated within the MSSM itself and
the assumption of GUT is not necessarily required. The results in both
cases are indeed impressive as we have discussed.  Certainly one can
add some more comments on the Finite Unified Theories. These are
related to some fundamental developments in Theoretical Particle
Physics based on reconsiderations of the problem of divergencies and
serious attempts to solve it. They include the motivation and
construction of string and noncommutative theories, as well as $N = 4$
supersymmetric field theories \cite{Mandelstam:1982cb,Brink:1982wv},
$N = 8$ supergravity
\cite{Bern:2009kd,Kallosh:2009jb,Bern:2007hh,Bern:2006kd,Green:2006yu}
and the AdS/CFT correspondence \cite{Maldacena:1997re}. It is a
thoroughly fascinating fact that many interesting ideas that have
survived various theoretical and phenomenological tests, as well as
the solution to the UV divergencies problem, find a common ground in
the framework of $N = 1$ Finite Unified Theories, which have been
discussed here. From the theoretical side they solve the problem of UV
divergencies in a minimal way.  On the phenomenological side in both
cases of reduction of couplings discussed here the celebrated success
of predicting the top-quark mass
\cite{Kapetanakis:1992vx,Kubo:1994bj} 
is now extended to the predictions of the Higgs masses and the
supersymmetric spectrum of the MSSM, which so far have been confronted
very successfully with the findings and bounds at the LHC.

The various scenarios will be refined/scrutinized in various ways in the
upcoming years. Important improvements in the analysis are expected from
progress on the theory side, in particular in an improved calculation of the
light Higgs boson mass. The corrections introduced in \cite{Hahn:2013ria} not only introduce
a shift in $\Mh$ (which should to some extent be covered by the estimate of
theory uncertainties). They will also reduce the theory uncertainties, see
\cite{Hahn:2013ria,Buchmueller:2013psa}, and in this way refine the selected model points, leading to a sharper
prediction of the allowed spectrum. One can hope that with even more
higher-order corrections included in the $\Mh$ calculation an uncertainty
below the $0.5 \gev$ level can be reached. 

The other important improvements in the future will be the continuing searches
for SUSY particles at collider experiments. The LHC will re-start in 2015 with
an increased center-of-mass enery of $\sqrt{s} \lsim 14 \tev$, largely
extending its SUSY search reach. The lower parts of the currently
allowed/predicted colored SUSY spectra will be tested in this way. For the
electroweak particles, on the other hand, $e^+e^-$ colliders might be the
better option. The ILC, operating at $\sqrt{s} \lsim 1 \tev$, has only a
limited potential for our model spectra. Going to higher energies, 
$\sqrt{s} \lsim 3 \tev$, that might be realized at CLIC, large parts of the
predicted electroweak model spectra can be covered. 

All spectra, however, (at least with the current calculation of $\Mh$
and its corresponding uncertainty), contain parameter regions that will escape
the searches at the LHC, the ILC and CLIC. In this case we would
remain with a light Higgs boson in the decoupling limit, i.e.\ would
be undistinguishable from a SM Higgs boson. The only hope to overcome
this situation is that an improved $\Mh$ calculation would cut away
such high spectra.

\section*{Acknowledgements}
G.Z. thanks the Institut f\"ur Theoretische Physik, Heidelberg, for
its generous support and warm hospitality.  The work of G.Z. was
supported by the Research Funding Program ARISTEIA, Higher Order
Calculations and Tools for High Energy Colliders, HOCTools
(co-financed by the European Union (European Social Fund ESF) and
Greek national funds through the Operational Program Education and
Lifelong Learning of the National Strategic Reference Framework
(NSRF)). G.Z and N.T. acknowledge also support from the European
Union's ITN programme HIGGSTOOLS.  The work of M.M. was supported by
mexican grants PAPIIT IN113712 and Conacyt 132059. The work of S.H.\
was supported in part by CICYT (grant FPA 2010--22163-C02-01) and by
the Spanish MICINN's Consolider-Ingenio 2010 Program under grant
MultiDark CSD2009-00064.  

%

\end{document}